# Interdisciplinarity metric based on the co-citation network

**Juan María Hernández [1] and Pablo Dorta-González [2,\***

[1] University of Las Palmas de Gran Canaria; juan.hernandez@ulpgc.es
[2] University of Las Palmas de Gran Canaria; pablo.dorta@ulpgc.es
**\*** Correspondence: pablo.dorta@ulpgc.es

**Abstract:** Quantifying the interdisciplinarity of a research is a relevant problem in the evaluative bibliometrics. The concept of interdisciplinarity is ambiguous and multidimensional. Thus, different measures of interdisciplinarity have been propose in the literature. However, few studies have proposed interdisciplinary metrics without previously defining classification sets, and no one use the co-citation network for this purpose. In this study we propose an interdisciplinary metric based on the co-citation network. This is a way to define the publication's field without resorting to pre-defined classification sets. We present a characterization of a publication's field and then we use this definition to propose a new metric of the interdisciplinarity degree for publications (papers) and journals as units of analysis. The proposed measure has an aggregative property that makes it scalable from a paper individually to a set of them (journal) without more than adding the numerators and denominators in the proportions that define this new indicator. Moreover, the aggregated value of two or more units is strictly among all the individual values.

## 1. Introduction

There is no consensus in the literature about the definition of *interdisciplinary research* (IDR) [1,2]. As a consequence, numerous indicators only try to measure one of its dimensions. The concept of interdisciplinarity is related to academic disciplines as a synthesis of theories and methods. However, there is considerable ambiguity with the discipline concept and its delimitation [3]. Historically, disciplines have been associated to the organization of teaching at universities. Nevertheless, nowadays the concept has become more general and also includes the creating new knowledge [4]. Focusing on knowledge creation, Sugimoto and Weingart [3] claim that IDR can be analysed from the dimensions publications, people, and ideas. These three dimensions can be measure through information from publications in multidisciplinary bibliographical databases.

In the scientometric approaches, most measures of interdisciplinarity are based on disciplinary delineations respect to indexing and classification of publications and/or their journals, mainly based on the publication perspective suggested by Sugimoto and Weingart [3]. However, with no conceptual consensus and plenty of dimensions, these IDR measures based on scientometric techniques has been interpreted in different ways [5,6]. Moreover, the choice of different classification sets and methodologies produces inconsistent and sometimes contradictory results [7]. Therefore, the current measurements of interdisciplinarity should be interpreted with caution in evaluative studies and science policies [8].

As indicated, numerous metrics have been propose for measuring interdisciplinarity, but only a few of them have used the network of citations, and no one the co-citation network. The aim of the present study is to define the unit's field without resorting to pre-defined classification systems. For this purpose we use the *co-citation network*. To define the field of a focal publication *i*, all publications co-cited with *i* are recognized. A publication *j* is co-cited with *i* if there is a third publication in which *i* and *j* are both cited. The publications co-cited with *i* are used to define the field of publication *i*.

Then we propose a new measure for the degree of interdisciplinarity and we analyse its properties. The aggregative property makes it scalable from a paper individually to a set of them (journal) without more than adding the numerators and denominators in the proportions that define the metric. Moreover, the aggregated value of the metric for two or more publications is strictly between the minimum and the maximum values of the metric for each one of them.

## 2. Interdisciplinary metrics based on the citation network

We focus our overview to interdisciplinarity metrics based on the publication dimension on networks (i.e., publications and their citation links) and the studies of inconsistent and non-robustness. The nodes in a citation network are formed by some papers and those other papers cited by them, and the edges between the nodes mean a citation link (see Figure 1, left).

To measure the degree of interdisciplinarity of journals, Leydesdorff [9] proposes the *betweenness-centrality* (BC) index. BC measures the degree of centrality for a node located on the shortest path between two other nodes in a network [10]. If a journal or a *subject category* (SC) is in betweenness other journals or SCs, its publications function as a communication channel for others and can be considered as interdisciplinary [11]. Recently, Leydesdorff et al. [12] modify the Rao-Stirling diversity and found this new indicator correlates with BC significantly more than Rao-Stirling diversity.

Rafols et al. [2] propose a *cluster coefficient* (CC) for the degree of interdisciplinarity of a SC. They identify the proportion of references among SCs, and then weighted by the percentage of publications that each SC has over the total number of publications. However, previous measures of IDR are inconsistent and non-robust. These measures may be problematic when used in practice because the IDR is strongly dependent on the chosen measure [8]. Furthermore, the choice of data and methodology can produce seriously inconsistent results [7]. Then, the metrics of interdisciplinarity should be used wisely in evaluative studies and science policies [8].

This inconsistent and non-robustness of the IDR metrics based on the citation network has motivated us to propose a new methodology based on the co-citation network in the following section.

## 3. Characterization of a publication's field though the co-citation network: An interdisciplinarity metric

Figure 1 illustrates a simple example of the citation and co-citation network of a paper $i$ (the focal paper). On the left hand side it is the citation network, where $s_i$ denotes the paper $i$'s degree or, in other words, paper $i$'s number of cites. Identically it is defined $s_j$, with $j = \{i_1, i_2, i_3\}$. The parameter $r_j$, with $j = \{i_1, i_2, i_3\}$ indicates the number of references of the $j$-th publication. On the right hand side it is the co-citation network, defined as the projection of the citation network on the set of cited papers. Parameter $k$ indicates the number of every paper's co-cites.

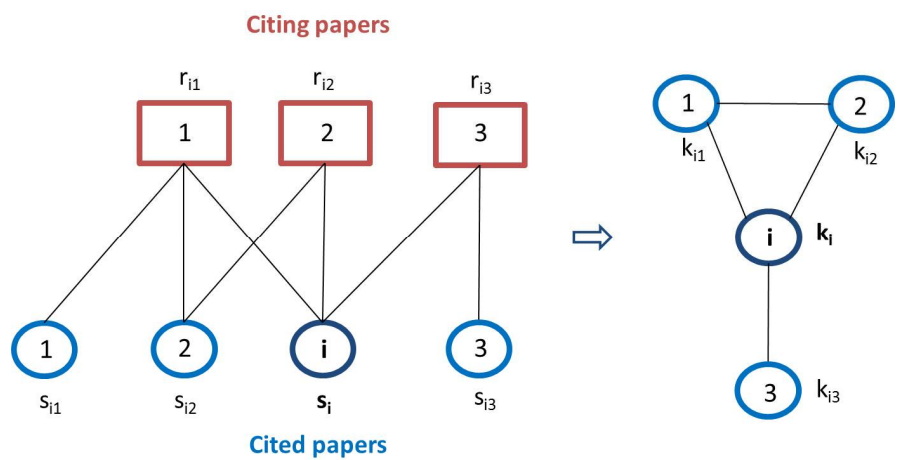

**Figure 1.** Citation (left) and co-citation (right) network of paper $i$. The square box indicate those papers citing $i$, while the circles show $i$'s co-cited papers. The parameter $s$ indicates the number of every paper's citations (cited paper 1 has $s_{i_1} = 1$, and similarly $s_{i_2} = 2$, $s_i = 3$, $s_{i_3} = 1$) and the parameter $r$ denotes the number of paper's references (citing paper 1 has $r_{i_1} = 3$ references, $r_{i_2} = 3$, $r_{i_3} = 2$). The parameter $k$ shows the number of every paper's co-citations ($k_{i_1} = 2$, $k_{i_2} = 2$, $k_i = 3$, $k_{i_3} = 1$)

We present a new metric, based on the relationship between the paper $i$'s degree in the citation ($s_i$) and co-citation network ($k_i$), which is

$$k_i = \sum_{j=1}^{s_i}(r_{i_j} - 1) - q_i. \tag{1}$$

The equation (1) shows that paper $i$'s degree in the co-citation network is equal to the sum of all references cited by the citing papers, excluding paper $i$ itself, and subtracting duplicated links $q_i$. A duplicated link is produced when two citing papers include paper $i$ and other paper $i'$ in the reference list. In this case, the connection between $i$ and $i'$ is duplicated. This is represented in Figure 1 by the X-motif between citing papers $\{1,2\}$ and cited papers $\{2, i\}$. Thus, $q_i$ represents the number of non-redundant X-motifs in the paper $i$'s citation network[1]. For example, $q_i = 1$ in Figure 1, and therefore $k_i = 2 + 1 + 1 - 1 = 3$.

From (1), we have the following equation,

$$\frac{k_i}{\sum_{j=1}^{s_i}(r_{i_j}-1)} = 1 - \frac{q_i}{\sum_{j=1}^{s_i}(r_{i_j}-1)}. \tag{2}$$

Then, we can define a new metric for paper $i$, starting from the following quotient,

$$XM_i = \frac{q_i}{\sum_{j=1}^{s_i}(r_{i_j}-1)}. \tag{3}$$

This parameter presents values in $[0,1)$ and indicates the proportion of paper $i$'s duplicated co-citations (alternatively, $1 - XM_i$ indicates the proportion of non-duplicated co-citations). In order to compare measures among different papers, we need a normalized metric. To do this, we calculate the maximum value that this parameter can reach:

$$XM_i^* = \frac{(s_i-1)k_i}{s_i k_i} = \frac{s_i-1}{s_i}, \tag{4}$$

which is lower than one. Then, the normalized metric is

$$\overline{XM}_i = \frac{1}{XM_i^*} XM_i = \frac{s_i}{s_i-1} \frac{q_i}{\sum_{j=1}^{s_i}(r_{i_j}-1)}. \quad (5)$$

---

[1] Given two groups of papers $\{\{j_1, j_2\}, \{i_1, i_2\}\}$ and $\{\{j_1, j_3\}, \{i_1, i_2\}\}$ forming an X-motif, $\{\{j_2, j_3\}, \{i_1, i_2\}\}$ forms automatically an X-motif as well. We call the latter a redundant X-motif. We exclude them in the calculation of $q_i$.

Now, the metric has values in [0,1]. A $\overline{XM}_i$ value close to 1 indicates that the citing papers recurrently refer to this paper in a group formed by the same papers. Thus, $\overline{XM}_i$ value is an indicator of the paper insertion in a specific research area, defined here by those papers which are usually cited jointly. On the contrary, high values of $1-\overline{XM}_i$ indicate that paper $i$ is not usually referred with the same papers, what reveals that this paper cannot be inserted in a specific research area, being an interdisciplinary paper instead. Therefore, we define the ***Interdisciplinary Research Index*** as

$$IDRI_i = 1 - \overline{XM}_i. \quad (6)$$

To illustrate the metric, we apply it to paper $i$ in the citation network of Figure 1. We observe that there is one X-motif ($q_i = 1$) and three citing papers. We have that $XM_i = \frac{1}{2+1+1} = 0.25$. Therefore, 25% of the co-cited papers are duplicated. Normalizing, we have that $\overline{XM}_i = \frac{3}{2} 0.25 = 0.375$ and $IDRI_i = 1 - 0.375 = 0.625$. Thus, paper $i$ is $37.5\%$ representative of the field where it is inserted and has $62.5\%$ of interdisciplinarity.

$\overline{XM}_i$ can be also used to define specific research areas, formed by those papers/journals with high $\overline{XM}_i$ and cited together.

## 4. Extensions

### 4.1. $\overline{XM}$-metric for two papers jointly

Now we present an extension of the metric above to the case of two papers jointly.

Assume two papers $i$ and $i'$. Every paper has its own citation and co-citation network. Figure 2 presents a simple example of this case. The citation network of paper $i$ includes two X-motifs ($q_i = 2$), while the citation network of paper $i'$ includes only one X-motif ($q_{i'} = 1$). Both papers have one X-motif in common, the one formed by citing papers $\{2,4\}$ and cited papers $\{i, i'\}$.

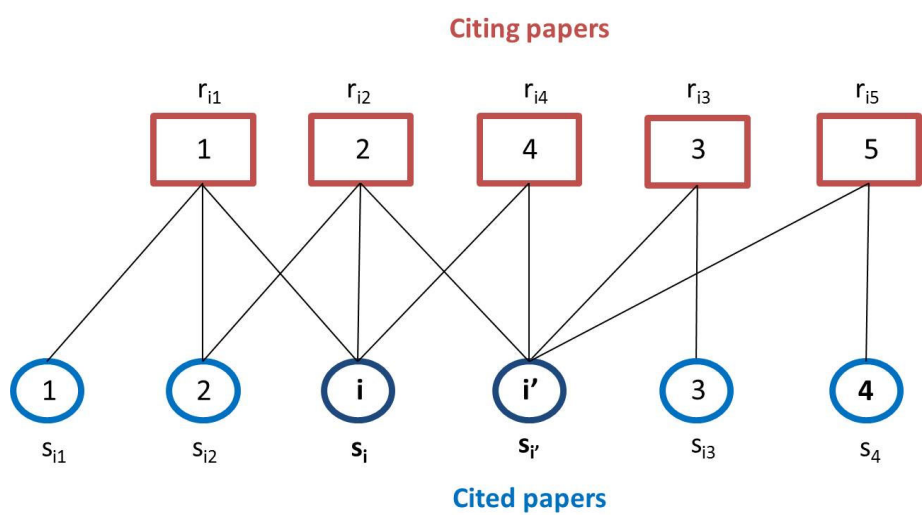

**Figure 2.** Citation network of papers $i$ and $i'$.

The XM-metric for the two papers jointly is built by simple aggregation of the number of X-motifs from the two papers. Thus, given the degree of papers $i$ and $i'$ in the citation network ($s_i$ and $s_{i'}$, respectively) and the number of non-redundant X-motifs for the papers $i$ and $i'$ ($q_i$ and $q_{i'}$, respectively), we define

$$XM_{\{i,i'\}} = \frac{q_i+q_{i'}}{\sum_{j=1}^{s_i}(r_{i_j}-1)+\sum_{j=1}^{s_{i'}}(r_{i'_j}-1)}. \quad (7)$$

This is the *mediant* calculation of $XM_i$ and $XM'_i$. In general, given non-negative real numbers $a, b, c, d$, with $bd \neq 0$, the mediant of the two fractions $\frac{a}{b}$ and $\frac{c}{d}$ is $\frac{a+c}{b+d}$. An important property is the *"mediant inequality"*, which indicates that the mediant lies strictly between the two fractions. Formally, if $\frac{a}{b} < \frac{c}{d}$ and $a, b, c, d > 0$, then $\frac{a}{b} < \frac{a+c}{b+d} < \frac{c}{d}$. This property follows from the two relations

$$\frac{a+c}{b+d} - \frac{a}{b} = \frac{bc-ad}{b(b+d)} = \frac{d}{b+d}\left(\frac{c}{d} - \frac{a}{b}\right) > 0, \quad (8)$$

$$\frac{c}{d} - \frac{a+c}{b+d} = \frac{bc-ad}{d(b+d)} = \frac{b}{b+d}\left(\frac{c}{d} - \frac{a}{b}\right) > 0. \quad (9)$$

Therefore, $XM_{\{i,i'\}}$ in (7) is in between the two fractions and coincides with the arithmetic mean if and only if the denominators of the fractions, reduced to the simplest form, are identical.

Parameter $XM_{\{i,i'\}}$ is again defined in the interval [0,1). The maximum value is the mediant of the maximum values of $XM_i$ and $XM_{i'}$. So, we define

$$\overline{XM}_{\{i,i'\}} = \frac{s_i+s_{i'}}{s_i+s_{i'}-2}\frac{q_i+q_{i'}}{\sum_{j=1}^{s_i}(r_{i_j}-1)+\sum_{j=1}^{s_{i'}}(r_{i'_j}-1)}. \quad (10)$$

This metric satisfies a nice property. We can assure that the $\overline{XM}$-metric for $\{i,i'\}$ is in between the $\overline{XM}$-metric for $i$ and $i'$. If this were not so, we can prove after some calculations that necessarily

$$(s_i - s_{i'})\left(\frac{q_i}{\sum_{j=1}^{s_i}(r_{i_j}-1)} - \frac{q_{i'}}{\sum_{j=1}^{s_{i'}}(r_{i'_j}-1)}\right) > 0. \quad (11)$$

In practice, it is expected that (11) is not fulfilled, since in co-citation networks where preferential attachment dominates [13], the relative number of X-motifs tends to zero as $s_i$ increases. This means that for large enough $s_i$, fraction $\frac{q_i}{\sum_{j=1}^{s_i}(r_{i_j}-1)}$ is a decreasing function of $s_i$, which is in contradiction to (11). Therefore, it is expected that the aggregated $\overline{XM}$-metric defined in (10) is in between the two $\overline{XM}$-metric for papers $i$ and $i'$.

We apply the metric to papers $\{i, i'\}$ in the citation network of Figure 2. Using the $\overline{XM}$-metric for single papers in (5), we have that $\overline{XM}_i = \frac{3}{2}\frac{2}{2+2+1} = 0.6$ and $\overline{XM}_{i'} = \frac{4}{3}\frac{1}{2+1+1+1} \simeq 0.27$. The $\overline{XM}$-metric for the two papers $i$ and $i'$ in common is $\overline{XM}_{\{i,i'\}} = \frac{3+4}{2+3}\frac{2+1}{5+5} = 0.42$. Thus, the set of papers $\{i, i'\}$ is 42.2% representative of the research area where it is inserted and has a 58% of interdisciplinarity.

### 4.2. $\overline{XM}$-metric for $n$ papers jointly

The generalization of the $\overline{XM}$-metric to $n$ papers ($n > 2$) is direct by applying the generalized mediant to $n$ fractions. Thus, given the set $\{i_1, i_2, \ldots, i_n\}$ of $n$ papers, the $\overline{XM}$-metric for these papers in common is defined as

$$\overline{XM}_{\{i_1,i_2,\ldots,i_n\}} = \frac{\sum_{k=1}^n s_{i_k}}{\sum_{k=1}^n s_{i_k}-n}\frac{\sum_{k=1}^n q_{i_k}}{\sum_{k=1}^n \sum_{j=1}^{s_{i_k}}(r_{i_k^j}-1)}, \quad (12)$$

where $r_{i_k^j}$ is the number of references of paper $j$, which is one of citing papers of $i_k$. Using the same argument above and proceeding by induction, it is expected that $\overline{XM}_{\{i_1,i_2,\ldots,i_n\}}$ is in between the two extreme values of $\overline{XM}$-metric for the $n$ papers.

## 5. Empirical application

In this cases study, 30 research articles published in 2018 are considered. As source of citations, the Scopus database is used. For the generation of the co-citation network, all the citations received for each of the 30 papers (up to the moment of this application, March 20, 2020) are considered.

These 30 papers are the most cited in five scientific journals (6 papers for each of the analyzed journals). The journals considered belong to the Library and Information Sciences subject category, and they were chosen trying to cover different sizes, according to the number of papers published in the year 2018, and different impact factors. This justifies that while one of the papers has cited 68 times so far, another has cited only 2 times. The aim of this is to show the application of the interdisciplinarity metric regardless of the number of citations available to generate the co-citation network. Notice that even with only two citations, it is already possible to generate the co-citation network.

The metadata for the identification of each paper (authors, journal, volume, number, and pages) are shown in Table 1. This table also includes the number of citations received, the number of nodes in the co-citation network, and the interdisciplinarity metric in percentage (IDRI x 100%) to facilitate its interpretation.

**Table 1.** Cases study with 30 research articles from six journals, with different size and impact factor, in the Library and Information Sciences subject category (Source of citations: Scopus)

| Authors | Journal Metadata | Citations | Nodes in co-citation network | IDRI x 100% |
|---|---|---|---|---|
| Abramo (2018) | Journal of Informetrics 12(3), 590-597 | 18 | 748 | 88.1% |
| Aslam (2018) | Library Management 39(1-2), 78-92 | 6 | 289 | 85.2% |
| Bates (2018) | Journal of Documentation 74(2), 412-429 | 10 | 593 | 95.8% |
| Bawden & Robinson (2018) | Journal of Documentation 74(1), 2-17 | 8 | 729 | 91.9% |
| Boyack et al. (2018) | Journal of Informetrics 12(1), 59-73 | 41 | 1,095 | 71.6% |
| Buschman (2018) | Library Quarterly 88(1), 23-40 | 7 | 456 | 93.1% |
| Clarke (2018) | Library Quarterly 88(1), 41-59 | 11 | 564 | 88.5% |
| Demir (2018) | Journal of Informetrics 12(4), 1296-1311 | 18 | 898 | 91.4% |
| Greifeneder et al. (2018) | Journal of Documentation 74(1), 119-136 | 13 | 633 | 91.9% |
| Hou et al. (2018) | Scientometrics 115(2), 869-892 | 25 | 1,609 | 91.5% |
| Javed & Liu (2018) | Scientometrics 115(1), 395-413 | 20 | 853 | 88.6% |
| Kulczycki et al. (2018) | Scientometrics 116(1), 463-486 | 36 | 945 | 77.1% |
| Lenstra (2018) | Library Quarterly 88(2), 142-159 | 3 | 172 | 94.5% |
| Leydesdorff et al. (2018) | Scientometrics 114(2), 567-592 | 19 | 733 | 83% |
| Li et al. (2018) | Scientometrics 115(1), 1-20 | 25 | 1,411 | 98.2% |
| Lor (2018) | Library Management 39(5), 307-321 | 4 | 142 | 88.2% |
| Martín-Martín et al. (2018) | Journal of Informetrics 12(4),1160-1177 | 68 | 4,723 | 94.1% |
| Mills et al. (2018) | Library Quarterly 88(2), 160-176 | 6 | 197 | 88.3% |
| Mwaniki (2018) | Library Management 39(1-2), 2-11 | 6 | 437 | 89.5% |
| Ocepek (2018) | Journal of Documentation 74(2), 398-411 | 8 | 369 | 85.2% |
| Orr (2018) | Library Quarterly 81(3), 399-423 | 3 | 2305 | 98.9% |
| Pan et al. (2018) | Journal of Informetrics 12(2), 481-493 | 22 | 1,451 | 93.8% |
| Ponelis & Adoma (2018) | Library Management 39(6-7), 430-448 | 2 | 69 | 92% |
| Rubenstein (2018) | Library Quarterly 88(2), 125-141 | 4 | 113 | 96.6% |
| Shepherd et al. (2018) | Library Management 39(8-9), 583-596 | 3 | 146 | 100% |
| Søe (2018) | Journal of Documentation 74(2), 309-332 | 8 | 333 | 91% |
| Spezi et al. (2018) | Journal of Documentation 74(1), 137-161 | 14 | 617 | 85.1% |
| Teixeira & Dobranszki (2018) | Scientometrics 115(2), 1107-1113 | 19 | 507 | 88.9% |
| Thelwall (2018) | Journal of Informetrics 12(2), 430-435 | 20 | 967 | 93% |
| Walter (2018) | Library Management 39(3-4), 154-165 | 4 | 266 | 99.2% |

As can be seen from the results obtained, the discipline considered is highly interdisciplinary, with values in the range from 71.6% to 100%. In 17 of the 30 cases, interdisciplinarity surpasses 90%. Half of the papers analyzed have an index higher than 91.5% (median) and the average is 90.5%.

As extreme cases in the range of variation, we comment two cases. The paper by Shepherd et al. (2018) has received three citations to date and the only common reference in these three documents is the mentioned one. This means that it has 100% variety in the co-citation network and its interdisciplinarity index is 1. On the contrary, the paper by Boyack et al. (2018) has received 41 citations to date, and among the references of said citing papers, a 71.6% of mismatched documents were found, representing an interdisciplinarity index of 0.716.

In this cases study, citations have not been limited to calendar years since the year of publication is quite recent in relation to the measurement of impact. The citation window could have been limited to the year 2019, but in that situation the number of citations would be less. However, in the case of analyzing authors instead of specific papers, this methodology allows adding all the papers of the same author, even from different years. If the analysis unit were the journal itself, it is also possible to add all the published documents in a specific year.

## 6. Conclusions

The relevance of the interdisciplinary research is well known. Many studies support its ability to solve complex problems and generate scientific developments and innovations [2,14]. As a consequence, funding agencies in many countries are considering the promotion of interdisciplinary research as a priority [15].

However, there is a lack of consistency and validity in the interdisciplinarity measures in the literature. The degree of interdisciplinarity varies with the selection of the metric, the source of the data, and the classification system used. Hence, different methodologies will produce different interdisciplinarity degrees [8]. Obviously, this generates a problem in research evaluation and science policy.

In this study we have proposed the co-citation network as a way to redefine the unit's field without resorting to a pre-defined classification system. The proposed new measure for the degree of interdisciplinarity is scalable from a unit individually (paper) to a set of them (journal), without more than adding the numerators and denominators in the proportions. Moreover, the aggregated value of two or more units is strictly among all the individual values. This important property of aggregability means that this new interdisciplinarity measure can also be applied at the meso (research groups, research centres) and macro levels (regions and countries). Note that as this metric is defined as a percentage, it is a relative value, so this indicator does not depend on the size of the unit of analysis [16]

An important application of this methodology could correspond to quantifying the scientific impact of publications. This problem is relevant when comparing the impact of publications from different scientific fields. This requires the use of metrics that normalize by the different citation habits between fields. [17]. In the citing-side normalization, each citation is weighted by the citation density of the citing field [18,19]. For cited-side normalization, this is the case of the Relative Citation Ratio (RCR), the articles co-cited with the focal article are utilized in the generation of the reference set that represents the field of the focal article [20]. However, this RCR has been criticized [21]. In this sense, we think our metric could be used in a new methodology for cited-side normalization.